\documentclass[a4paper,12pt]{article}
\usepackage{a4}
\usepackage{amsmath,amssymb}
\usepackage{color}
\usepackage{graphicx}
\usepackage{hyperref}

\newtheorem{proposition}{Proposition}

\newcommand{\Z}{\mathcal{Z}}
\newcommand{\norm}[1]{\frac{1}{c^{#1}}}
\newcommand{\vect}[1]{\underline{#1}}  
\newcommand{\abl}[2]{\frac{\partial #1}{\partial #2}} 
\newcommand{\set}{\stackrel{!}{=}}      

\newcommand{\mm}{\textsc{Mathematica$^\textrm{\textregistered}$ }}

\newcommand{\Blue}[1]{\textcolor{blue}{#1}}
\newcommand{\Green}[1]{\textcolor{green}{#1}}
\newcommand{\Red}[1]{\textcolor{red}{#1}}
\newcommand{\textBlue}[0]{\color{blue}}

\newcommand{\textBlack}[0]{\color{black}}

\title{Implementation of Liu's procedure in \mm for use in \\Relativistic Constitutive Theory}
\author{Heiko J. Herrmann}

\begin{document}
\bibliographystyle{unsrt}
\maketitle
\begin{abstract}\setlength{\parindent}{0em}
\noindent
The aim of this article is to show, how computer algebra can be used when applying Liu's procedure.\\
Although \mm (a commercial product by Wolfram Research Inc.) is used in this article, it is possible to use other computer algebra systems as well. 
\end{abstract}

\section{Thermodynamics}
\subsection{Liu's Procedure \\ (A \textit{very} short introduction)}
Here a \textit{very} short introduction will be presented, so that the reader will get a glance of what the Liu procedure is. For a detailed discussion of Liu's procedure and balance equations please consult the following literature: \cite{LIU72}, \cite{KIELCE96}, \cite{MU90}, \cite{MUPAEH01}, \cite{HHMUPARUE00},\cite{HHMUPARUE98}, \cite{MUEH96}, \cite{HH-DISS}
\begin{center}
\includegraphics[width=5cm]{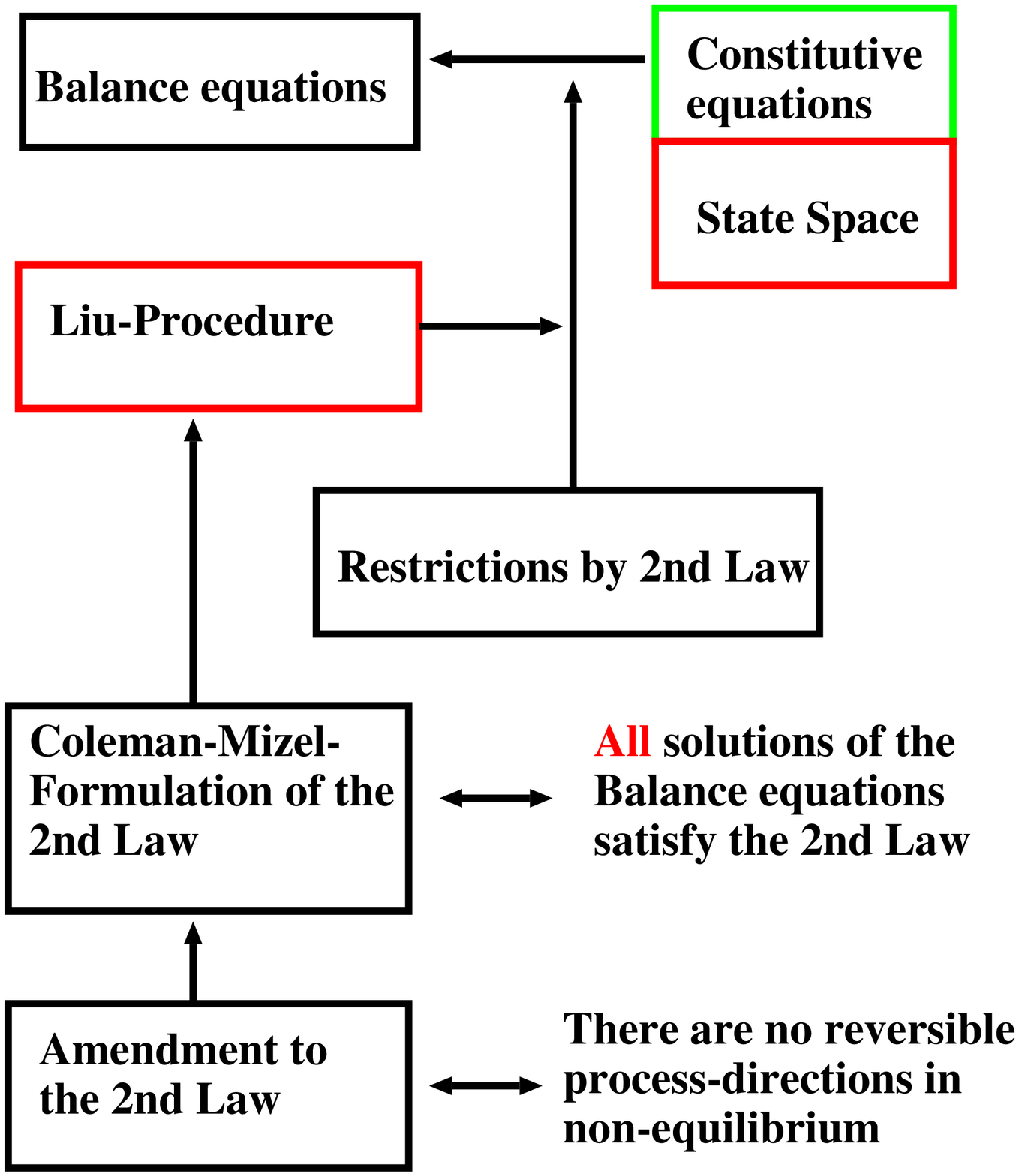}
\end{center}
\begin{proposition}[Coleman-Mizel formulation of the 2nd Law]\cite{COMI64}~\\
  If $\mathcal{Z}$ is no trap, then the following statement is valid for all $X$:
\begin{eqnarray}
A(\Z) \cdot X & = & - B(\Z) \\
\Rightarrow \alpha(\Z) \cdot X & \geq & - \beta(\Z)
\end{eqnarray}
For all $X$, which satisfy the balance equations, the dissipation inequality is fullfilled automatically.
\end{proposition}
\begin{proposition}[Liu proposition]\cite{LIU72}~\\
Starting with the Coleman-Mizel formulation of the Second Law one can show that in large state spaces there exist state functions so that the following relations are valid.
\begin{eqnarray*}
  \Lambda(\Z) \cdot A(\Z) & = & \alpha(\Z)\\
- \Lambda(\Z) \cdot B(\Z) & \geq & - \beta(\Z)
\end{eqnarray*}
\end{proposition}

This means that there exist restrictions to the constitutive equations.

\subsection{Example for the Application of Liu's Procedure\\ (for arbitrary non-relativistic balances)}
Here a short example for general balances and a state space is given \cite{HH-DISS}. This example shall demonstrate how the balances on the state space and the matrix formulation is achieved. Therefore two balances and an inequality for arbitrary quantities are considered.\\
The balances read:
\begin{eqnarray}
\partial_t a_1 + \nabla \cdot (\vect{v} a_1 + b_1) & = & c_1\\
\partial_t a_2 + \nabla \cdot (\vect{v} a_2 + b_2) & = & c_2\\
\partial_t \xi + \nabla \cdot (\vect{v} \xi + \gamma) & \geq & 0
\end{eqnarray}
The state space is
\begin{eqnarray}
\Blue{\Z} & = & \{ \Blue{a_1}, \Blue{a_2}, \Blue{\vect{v}}, \Blue{\partial_t a_1} \}
\end{eqnarray}
This is the domain of for the constitutive functions $b_1, b_2, \xi \gamma$.\\
Balances on the state space are obtained by using the chain rule when performing the derivatives.
\begin{eqnarray}
\Blue{\partial_t a_1} + \nabla \cdot (\Blue{\vect{v}} \Blue{a_1} + b_1(\Blue{\Z})) - c_1 & = & 0\\
\Blue{\partial_t a_1} + \Red{\nabla \cdot \vect{v}} \Green{a_1} + \Green{\vect{v}} \cdot \Red{\nabla a_1} + \nonumber \\
+ \Green{\abl{b_1}{a_1}} \cdot \Red{\nabla a_1} + \Green{\abl{b_1}{a_2}} \cdot \Red{\nabla a_2} + \Green{\abl{b_1}{\vect{v}}} \cdot \Red{\nabla \vect{v}} + \Green{\abl{b_1}{\partial_t a_1}} \cdot \Red{\nabla \partial_t a_1} - c_1 & = & 0
\end{eqnarray}
\begin{eqnarray}
\Red{\partial_t a_2} + \nabla \cdot (\Blue{\vect{v} a_2} + b_2(\Blue{\Z})) - c_2 & = & 0\\
\Red{\partial_t a_2} + \Red{\nabla \cdot \vect{v}} \Green{a_2} + \Green{\vect{v}} \cdot \Red{\nabla a_2} + \nonumber \\
+ \Green{\abl{b_2}{a_1}} \cdot \Red{\nabla a_1} + \Green{\abl{b_2}{a_2}} \cdot \Red{\nabla a_2} + \Green{\abl{b_2}{\vect{v}}} \cdot \Red{\nabla \vect{v}} + \Green{\abl{b_2}{\partial_t a_1}} \cdot \Red{\nabla \partial_t a_1} - c_2 & = & 0
\end{eqnarray}
\begin{eqnarray}
\partial_t \xi(\Blue{\Z}) + \nabla \cdot (\Blue{\vect{v}} \xi(\Blue{\Z}) + \gamma(\Blue{\Z})) & \geq & 0\\
\abl{\xi}{a_1} \cdot \Blue{\partial_t a_1} + \Green{\abl{\xi}{a_2}} \cdot \Red{\partial_t a_2} + \Green{\abl{\xi}{\vect{v}}} \cdot \Red{\partial_t \vect{v}} + \Green{\abl{\xi}{\partial_t a_1}} \cdot \Red{\partial_t \partial_t a_1} +\nonumber \\
+ \Red{\nabla \cdot \vect{v}} \Green{\xi} + \Green{\vect{v} \cdot \abl{\xi}{a_1}} \Red{\nabla a_1} + \Green{\vect{v} \cdot \abl{\xi}{a_2}} \cdot \Red{\nabla a_2} + \Green{\vect{v} \cdot \abl{\xi}{\vect{v}}} \cdot \Red{\nabla \vect{v}} + \Green{\vect{v} \cdot \abl{\xi}{\partial_t a_1}} \cdot \Red{\nabla \partial_t a_1} + \nonumber \\
+ \Green{\abl{\gamma}{a_1}} \cdot \Red{\nabla a_1} + \Green{\abl{\gamma}{a_2}} \cdot \Red{\nabla a_2} + \Green{\abl{\gamma}{\vect{v}}} \cdot \Red{\nabla \vect{v}} + \Green{\abl{\gamma}{\partial_t a_1}} \cdot \Red{\nabla \partial_t a_1} & \geq & 0
\end{eqnarray}
As $\Blue{\partial_t a_1}$ is included in the state space it is not a higher derivative.\\
 
Next the balances can be written in matrix formulation.
\begin{eqnarray}
\Green{\underline{\underline{A}}}(\Blue{\Z}) \cdot \Red{\vect{X}} + \vect{B}(\Blue{\Z}) & = & 0\\
\Green{\underline{\alpha}}(\Blue{\Z}) \cdot \Red{\vect{X}} + \beta(\Blue{\Z}) & \geq & 0
\end{eqnarray}
 
\begin{eqnarray}
  \Red{((X))}^\top & = & \left( \begin{array}{ccccccc} \Red{\partial_t a_2} & \Red{\partial_t \vect{v}} & \Red{\partial_t \partial_t a_1} & \Red{\nabla a_1} & \Red{\nabla a_2} & \Red{\nabla \vect{v}} & \Red{\nabla \partial_t a_1} \end{array} \right)\\
~\nonumber\\
\Green{((A))} & = & \left(
\begin{array}{ccccccc}
0 & 0 & 0 & \Green{\vect{v}} + \Green{\abl{b_1}{a_1}} & \Green{\abl{b_1}{a_2}} & \Green{a_1 \vect{\vect{1}}} + \Green{\abl{b_1}{\vect{v}}} & \Green{\abl{b_1}{\partial_t a_1}}\\
~\\
 1 & 0 & 0 & \Green{\abl{b_2}{a_1}} & \Green{\vect{v}} + \Green{\abl{b_2}{a_2}} & \Green{a_2 \vect{\vect{1}}} + \Green{\abl{b_2}{\vect{v}}} & \Green{\abl{b_2}{\partial_t a_1}}
\end{array} \right)\\
~\nonumber\\
((B)) & = & \left(
\begin{array}{c}
\Blue{\partial_t a_1} - c_1\\
- c_2
\end{array} \right)\\
~\nonumber\\
\Green{((\alpha))} & = & \left(
\begin{array}{ccccccc}
\Green{\abl{\xi}{a_2}} & \Green{\abl{\xi}{\vect{v}}} & \Green{\abl{\xi}{\partial_t a_1}} & \Green{\vect{v} \cdot \abl{\xi}{a_1}} + \Green{\abl{\gamma}{a_1}} & \Green{\vect{v} \cdot \abl{\xi}{a_2}} + \Green{\abl{\gamma}{a_2}} & \Green{\xi \vect{\vect{1}}} + \Green{\vect{v} \cdot \abl{\xi}{\vect{v}}} + \Green{\abl{\gamma}{\vect{v}}} & \Green{\vect{v} \cdot \abl{\xi}{\partial_t a_1}} + \Green{\abl{\gamma}{\partial_t a_1}}
\end{array} \right) \nonumber\\
~\\
~\nonumber\\
((\beta)) & = & \left(
\begin{array}{c}
\abl{\xi}{a_1} \cdot \partial_t a_1
\end{array} \right)
\end{eqnarray}
Next the Liu equations can be written down
\begin{eqnarray}
\vect{\Lambda}(\Blue{\Z}) \cdot \Green{\underline{\underline{A}}}(\Blue{\Z}) & = & \Green{\vect{\alpha}}(\Blue{\Z})\\
- \vect{\Lambda}(\Blue{\Z}) \cdot \vect{B}(\Blue{\Z}) & \geq & - \beta(\Blue{\Z})
\end{eqnarray}
\begin{eqnarray}
\Lambda_2 & = & \Green{\abl{\xi}{a_2}}\\
0 & = & \Green{\abl{\xi}{\vect{v}}}\\
0 & = & \Green{\abl{\xi}{\partial_t a_1}}\\
\Lambda_1 \left(\Green{\vect{v}} + \Green{\abl{b_1}{a_1}} \right) + \Lambda_2 \left(\Green{\abl{b_2}{a_1}}\right) & = & \Green{\vect{v} \cdot \abl{\xi}{a_1}} + \Green{\abl{\gamma}{a_1}}\\
\Lambda_1 \Green{\abl{b_1}{a_2}} + \Lambda_2 \left( \Green{\vect{v}} + \Green{\abl{b_2}{a_2}}\right) & = & \Green{\vect{v} \cdot \abl{\xi}{a_2}} + \Green{\abl{\gamma}{a_2}}\\
\Lambda_1 \left(\Green{a_1 \vect{\vect{1}}} + \Green{\abl{b_1}{\vect{v}}}\right) + \Lambda_2 \left( \Green{a_2 \vect{\vect{1}}} + \Green{\abl{b_2}{\vect{v}}} \right) & = & \Green{\xi} + \Green{\vect{v} \cdot \abl{\xi}{\vect{v}}} + \Green{\abl{\gamma}{\vect{v}}}\\
\Lambda_1 \Green{\abl{b_1}{\partial_t a_1}} + \Lambda_2 \Green{\abl{b_2}{\partial_t a_1}} & = & \Green{\vect{v} \cdot \abl{\xi}{\partial_t a_1}} + \Green{\abl{\gamma}{\partial_t a_1}}
\end{eqnarray}
The residual inequality reads
\begin{eqnarray}
\Lambda_1 (\Blue{\partial_t a_1} - c_1 ) - \Lambda_2 c_2 & \leq & \abl{\xi}{a_1} \cdot \partial_t a_1
\end{eqnarray}
Because of the principle of \textit{material frame indifference} all derivatives with respect to $\vect{v}$ shall be set equal to zero: $\abl{}{\vect{v}} \set 0$. The \textit{material frame indifference} means that material laws shall be independent of the velocity of the observer. Using this the Liu equations read:
\begin{eqnarray}
\Lambda_2 & = & \Green{\abl{\xi}{a_2}}\\
0 & = & \Green{\abl{\xi}{\vect{v}}}\\
0 & = & \Green{\abl{\xi}{\partial_t a_1}}\\
\Lambda_1 \left(\Green{\vect{v}} + \Green{\abl{b_1}{a_1}} \right) + \Lambda_2 \left(\Green{\abl{b_2}{a_1}}\right) & = & \Green{\vect{v} \cdot \abl{\xi}{a_1}} + \Green{\abl{\gamma}{a_1}}\\
\Lambda_1 \Green{\abl{b_1}{a_2}} + \Lambda_2 \left( \Green{\vect{v}} + \Green{\abl{b_2}{a_2}}\right) & = & \Green{\vect{v} \cdot \abl{\xi}{a_2}} + \Green{\abl{\gamma}{a_2}}\\
\Lambda_1 \left( \Green{a_1 \vect{\vect{1}}} \right) + \Lambda_2 \left( \Green{a_2 \vect{\vect{1}}} \right) & = & \Green{\xi} \\
\Lambda_1 \Green{\abl{b_1}{\partial_t a_1}} + \Lambda_2 \Green{\abl{b_2}{\partial_t a_1}} & = & \Green{\vect{v} \cdot \abl{\xi}{\partial_t a_1}} + \Green{\abl{\gamma}{\partial_t a_1}}
\end{eqnarray}
The residual inequality remains unchanged
\begin{eqnarray}
\Lambda_1 (\Blue{\partial_t a_1} - c_1 ) - \Lambda_2 c_2 & \leq & \abl{\xi}{a_1} \cdot \partial_t a_1
\end{eqnarray}

From these equations the Lagrange parameters $\underline{\Lambda}(\Blue{\Z})$ can be determined.
\begin{eqnarray}
\Lambda_2 & = & \Green{\abl{\xi}{a_2}}\\
\Lambda_1 & = & \left( \Green{\xi} -\Green{\abl{\xi}{a_2}} \left( \Green{a_2 \vect{\vect{1}}} \right) \right) \left( \Green{a_1 \vect{\vect{1}}} \right)^{-1}
\end{eqnarray}
 
After inserting the Lagrange parameters into Liu's equations the restrictions to constitutive equations are given
\begin{eqnarray}
0 & = & \Green{\abl{\xi}{\vect{v}}}\\
0 & = & \Green{\abl{\xi}{\partial_t a_1}}\\
\left( \Green{\xi} -\Green{\abl{\xi}{a_2}} \left( \Green{a_2 \vect{\vect{1}}} \right) \right) \left( \Green{a_1 \vect{\vect{1}}} \right)^{-1} \left(\Green{\vect{v}} + \Green{\abl{b_1}{a_1}} \right) + \Green{\abl{\xi}{a_2}} \left(\Green{\abl{b_2}{a_1}}\right) & = & \Green{\vect{v} \cdot \abl{\xi}{a_1}} + \Green{\abl{\gamma}{a_1}}\\
\left( \Green{\xi} -\Green{\abl{\xi}{a_2}} \left( \Green{a_2 \vect{\vect{1}}} \right) \right) \left( \Green{a_1 \vect{\vect{1}}} \right)^{-1} \Green{\abl{b_1}{a_2}} + \Green{\abl{\xi}{a_2}} \left( \Green{\vect{v}} + \Green{\abl{b_2}{a_2}}\right) & = & \Green{\vect{v} \cdot \abl{\xi}{a_2}} + \Green{\abl{\gamma}{a_2}}\\
\left( \Green{\xi} -\Green{\abl{\xi}{a_2}} \left( \Green{a_2 \vect{\vect{1}}} \right) \right) \left( \Green{a_1 \vect{\vect{1}}} \right)^{-1} \Green{\abl{b_1}{\partial_t a_1}} + \Green{\abl{\xi}{a_2}} \Green{\abl{b_2}{\partial_t a_1}} & = & \Green{\vect{v} \cdot} \underbrace{\Green{\abl{\xi}{\partial_t a_1}}}_{= \ 0} + \Green{\abl{\gamma}{\partial_t a_1}}
\end{eqnarray}
The entropy production is given by the residual inequality
\begin{eqnarray}
\sigma \ := \ - \left(\Green{\vect{v} \cdot \abl{\xi}{a_2}} \right) \left( \Green{\abl{b_1}{a_2}} \right)^{-1} (\Blue{\partial_t a_1} - c_1 ) + \abl{\xi}{a_1} & \geq & 0
\end{eqnarray}
 
~\\
This example was meant to demonstrate how the Liu equations and the residual inequality are achieved. This example shows also that the entropy production is independent of the highest derivatives included in the state space (this is a general property of the entropy density; see also \cite{MUDO96}). \\

\section{Application of Liu's Procedure using Mathematica}
For the correct handling of indices I will use the package ''Ricci.m'' by John M. Lee, which can be found on the MathSource server.
\textBlue \begin{verbatim}
<<Ricci`
\end{verbatim} \textBlack

\subsection{Declaration of Tensors and State Space}
First it is necessary to define a bundle and declare the letters used for indices.
\textBlue \begin{verbatim}
DefineBundle[fiber2,4,g,{i,j,k,l,m,n,o,r,y}]
\end{verbatim} \textBlack
Next the tensors have to be defined. First I will declare the state space variables (z0, \dots, z5).
\begin{eqnarray*}
  \Z & = & \{ n0, u, p, e, spa, spv \}
\end{eqnarray*}
\textBlue \begin{verbatim}
(* z0=n0 *) DefineTensor[n0,0] (* particle number density *)
(* z1 =u *) DefineTensor[u,1] (* four velocity *)
(* z2=p *) DefineTensor[p,1] (* momentum flux *)
(* z3=e *) DefineTensor[e,0] (* energy *)
(* z4=spa *) DefineTensor[spa,2, Symmetries -> Skew] (* spin density *)
(* z5=spv *) DefineTensor[spv,1] (* vector spin density *)
\end{verbatim} \textBlack
Then the constitutive functions are declared:
\textBlue \begin{verbatim}
DefineTensor[t,2] (* stress tensor *)
DefineTensor[q,1] (* heat flux *)
DefineTensor[scs,3] (* couple stress *)
DefineTensor[spb,2] (* 2-couple stress *)
DefineTensor[s,0] (* entropy *)
DefineTensor[se,1] (* entropy flux *)
DefineTensor[f,1] (* external force *)
\end{verbatim} \textBlack

\subsection{Balance Equations}
The next step is to write down the balance equations, which are for technical reasons resolved to zero, and the right hand side ($ =0 $) is not written down:\\
Balance of particle number density:
\begin{eqnarray*}
  \partial_l \left(n u^l\right) & = & 0
\end{eqnarray*}
\textBlue \begin{verbatim}
D[n0[x] u[U[l]][x],x])
\end{verbatim} \textBlack
Balance of energy momentum
\begin{eqnarray*}
  \partial_l (t^{il} + \norm{2} p^i u^l + \norm{2} u^i q^l + \norm{4} e u^i u^l) -f^i & = & 0
\end{eqnarray*}
\textBlue \begin{verbatim}
D[t[U[i],U[l]][x]+1/c^2  p[U[i]][x]u[U[l]][x] + 1/c^2 
        u[U[i]][x]q[U[l]][x]+1/c^4 e[x]u[U[i]][x]u[U[l]][x],x ] - f[U[i]]
\end{verbatim} \textBlack
Balance of spin
\begin{eqnarray*}
  \partial_l \left(scs^{ijl} + \norm{2} spa^{ij} u^l + \norm{2} u^{[i} spb^{j]l} + \norm{4} u^{[i}spv^{j]} u^l \right) - \left(t^{[ij]} + \norm{2} p^{[i} u^{j]} + \norm{2} u^{[i} q^{j]}\right) & = & 0
\end{eqnarray*}
\textBlue \begin{verbatim}
D[scs[U[i],U[j],U[l]][x] + 
      1/c^2 spa[U[i],U[j]] u[U[l]] + 1/(2  c^2)u[U[i]] spb[U[j],U[l]][x] - 
      1/(2 c^2) u[U[j]] spb[U[i],U[l]][x] + 
      1/(2 c^4) u[U[i]] spv[U[j]] u[U[l]] - 
      1/(2 c^4) u[U[j]] spv[U[i]] u[U[l]],x] - 
  1/2 (t[U[i],U[j]] + 1/c^2 p[U[i]] u[U[j]] + 1/c^2 u[U[i]]q[U[j]] - 
        t[U[j],U[i]] + 1/c^2 p[U[j]] u[U[i]] + 1/c^2 u[U[j]]q[U[i]])
\end{verbatim} \textBlack
Balance of entropy
\begin{eqnarray*}
  \partial_l \left(s u^l + \norm{2} se^l\right) & \geq & 0
\end{eqnarray*}
\textBlue \begin{verbatim}
D[s [x]u[U[l]][x] + 1/c^2 se[U[l]][x],x]
\end{verbatim} \textBlack

\newpage
\subsection{Balances on State Space}
As the constitutive functions are defined on the state space, the derivatives have to be performed by use of the chain rule. The equations are ''labeled'' for further use.
\begin{eqnarray*}
  \partial_l \left(n u^l\right) & = & 0
\end{eqnarray*}
\textBlue \begin{verbatim}
TB = D[n0[x] u[U[l]][x],x]
\end{verbatim} \textBlack
\begin{eqnarray*}
  \partial_l \left(t^{il}(\Z) + \norm{2} p^i u^l + \norm{2} u^i q^l(\Z) + \norm{4} e u^i u^l \right) - f^i & = & 0
\end{eqnarray*}
\textBlue \begin{verbatim}
ImpB=\[Kappa] 
      D[t[U[i],U[l]][z0[x],z1[x],z2[x],z3[x],z4[x],z5[x]] + 
            1/c^2  p[U[i]][x]u[U[l]][x] +
            1/c^2  u[U[i]][x]q[U[l]][z0[x],z1[x],z2[x],z3[x],z4[x],z5[x]] +
            1/c^4 e[x]u[U[i]][x]u[U[l]][x],x] - f[U[i]]
\end{verbatim} \textBlack
\begin{eqnarray*}
  \partial_l \left(scs^{ijl}(\Z) + \norm{2} u^{[i} spb^{j]l}(\Z) + \norm{4} u^{[i}spv^{j]} u^l \right) - \left(t^{[ij]} + \norm{2} p^{[i} u^{j]} + \norm{2} u^{[i} q^{j]}\right) & = & 0
\end{eqnarray*}
\textBlue \begin{verbatim}
SpinB=D[scs[U[i],U[j],U[l]][z0[x],z1[x],z2[x],z3[x],z4[x],z5[x]] + 
        1/c^2 spa[U[i],U[j]] u[U[l]] +
        1/(2  c^2)u[U[i]] spb[U[j],U[l]][z0[x],z1[x],z2[x],z3[x],z4[x],z5[x]] -
        1/(2 c^2) u[U[j]] spb[U[i],U[l]][z0[x],z1[x],z2[x],z3[x],z4[x],z5[x]] +
        1/(2 c^4) u[U[i]] spv[U[j]] u[U[l]] - 
        1/(2 c^4) u[U[j]] spv[U[i]] u[U[l]],x] - 
    1/2 (t[U[i],U[j]] + 1/c^2 p[U[i]] u[U[j]] + 1/c^2 u[U[i]]q[U[j]] - 
          t[U[j],U[i]] + 1/c^2 p[U[j]] u[U[i]] + 1/c^2 u[U[j]]q[U[i]])
\end{verbatim} \textBlack
\begin{eqnarray*}
\partial_l \left(s(\Z) u^l + \norm{2} se^l(\Z) \right) & \geq & 0
\end{eqnarray*}
\textBlue \begin{verbatim}
EntropieB=
  D[s [z0[x],z1[x],z2[x],z3[x],z4[x],z5[x]]u[U[l]][x] + 
    1/c^2 se[U[l]][z0[x],z1[x],z2[x],z3[x],z4[x],z5[x]],x]
\end{verbatim} \textBlack

\newpage
\textBlack
\subsection{Matrix Formulation}
Now the components of the matrices for the matrix formulation of the balances have to be collected.\\
Matrix of higher derivatives
\textBlue \begin{verbatim}
X0:=z0'[x]
X1:=z1'[x]
X2:=z2'[x]
X3:=z3'[x]
X4:=z4'[x]
X5:=z5'[x]
\end{verbatim} \textBlack
Then the higher derivatives have to be set correctly by hand.
\textBlue \begin{verbatim}
(z0)'[x_]:=n0[L[l]]
(z1)'[x_]:=u[U[n]][L[l]] 
(z2)'[x_]:=p[U[n]][L[l]] 
(z3)'[x_]:=e[L[l]]
(z4)'[x_]:=spa[U[n],U[m]][L[l]]
(z5)'[x_]:=spv[U[n]][L[l]]
(n0)'[x_]:=n0[L[l]]
(u[U[i]])'[x_]:=u[U[i]][L[l]]
(u[U[j]])'[x_]:=u[U[j]][L[l]]
(u[U[l]])'[x_]:=u[U[l]][L[l]]
(p[U[i]])'[x_]:=p[U[i]][L[l]]
(p[U[j]])'[x_]:=p[U[j]][L[l]]
(e)'[x_]:=e[L[l]]
\end{verbatim} \textBlack
Then the coefficient matrix
\textBlue \begin{verbatim}
A00=Coefficient[TB,z0'[x]]
A01=Coefficient[TB,z1'[x]]+g[L[n],U[l]] Coefficient[TB,(u[U[l]])'[x]]
A02=Coefficient[TB,z2'[x]]
A03=Coefficient[TB,z3'[x]]
A04=Coefficient[TB,z4'[x]]
A05=Coefficient[TB,z5'[x]]
A10=Coefficient[ImpB,z0'[x]]
A11=Coefficient[ImpB,z1'[x]]+g[L[n],U[i]] Coefficient[ImpB,(u[U[i]])'[x]]+
    g[L[n],U[l]] Coefficient[ImpB,(u[U[l]])'[x]]
A12=Coefficient[ImpB,z2'[x]]+g[L[n],U[i]] Coefficient[ImpB,(p[U[i]])'[x]]
A13=Coefficient[ImpB,z3'[x]]
A14=Coefficient[ImpB,z4'[x]]
A15=Coefficient[ImpB,z5'[x]]
A20=Coefficient[SpinB,z0'[x]]
A21=Coefficient[SpinB,z1'[x]]+g[L[n],U[i]] Coefficient[SpinB,(u[U[i]])'[x]] +
    g[L[n],U[j]] Coefficient[SpinB,(u[U[j]])'[x]]+
    g[L[n],U[l]] Coefficient[SpinB,(u[U[l]])'[x]]
A22=Coefficient[SpinB,z2'[x]]+g[L[n],U[i]] Coefficient[SpinB,(p[U[i]])'[x]]+
    g[L[n],U[j]] Coefficient[SpinB,(p[U[j]])'[x]]
A23=Coefficient[SpinB,z3'[x]]
A24=Coefficient[SpinB,z4'[x]]
A25=Coefficient[SpinB,z5'[x]]
\end{verbatim} \textBlack
Coefficient matrix of dessipation inequality
\textBlue \begin{verbatim}
\[Alpha]0=Coefficient[EntropieB,z0'[x]]
\[Alpha]1=
  Coefficient[EntropieB,z1'[x]]+
    g[L[n],U[l]] Coefficient[EntropieB,(u[U[l]])'[x]]
\[Alpha]2=Coefficient[EntropieB,z2'[x]]
\[Alpha]3=Coefficient[EntropieB,z3'[x]]
\[Alpha]4=Coefficient[EntropieB,z4'[x]]
\[Alpha]5=Coefficient[EntropieB,z5'[x]]
\end{verbatim} \textBlack
The residual matrix
\textBlue \begin{verbatim}
B0=Simplify[
    TB - (A00 z0'[x] + A01  z1'[x]+A02  z2'[x]+A03  z3'[x]+A04  z4'[x]+
          A05  z5'[x])]
B1=Simplify[
    ImpB - (A10 z0'[x] + A11  z1'[x]+A12  z2'[x]+A13  z3'[x]+A14  z4'[x]+
          A15  z5'[x])]
B2= Simplify[
    SpinB - (A20 z0'[x] + A21  z1'[x]+A22  z2'[x]+A23  z3'[x]+A24  z4'[x]+
          A25  z5'[x])]
\end{verbatim} \textBlack
\textBlue \begin{verbatim}
\[Beta]= Simplify[
    EntropieB - (
        \[Alpha]0 z0'[x] + \[Alpha]1 z1'[x]+\[Alpha]2 z2'[x]+\[Alpha]3 z3'[x]+
          \[Alpha]4 z4'[x]+\[Alpha]5 z5'[x])]
\end{verbatim} \textBlack

\newpage
\subsection{Liu Relations}
Now Liu's proposition (based on Farkash's lemma) can be applied. The result are the Liu equations
\textBlue \begin{verbatim}
liu1= \[Lambda]0 A00 + \[Lambda]1 A10 + \[Lambda]2 A20 -\[Alpha]0
liu2= \[Lambda]0 A01 + \[Lambda]1 A11 + \[Lambda]2 A21 -\[Alpha]1
liu3= \[Lambda]0 A02 + \[Lambda]1 A12 + \[Lambda]2 A22 -\[Alpha]2
liu4= \[Lambda]0 A03 + \[Lambda]1 A13 + \[Lambda]2 A23 -\[Alpha]3
liu5= \[Lambda]0 A04 + \[Lambda]1 A14 + \[Lambda]2 A24 -\[Alpha]4
liu6= \[Lambda]0 A05 + \[Lambda]1 A15 + \[Lambda]2 A25 -\[Alpha]5
\end{verbatim} \textBlack
and the residual inequality
\textBlue \begin{verbatim}
\[Lambda]0 B0 + \[Lambda]1 B1 + \[Lambda]2 B2 -\[Beta]
\end{verbatim} \textBlack
The Lagrange parameters can be determined
\textBlue \begin{verbatim}
lsg=Solve[{liu1==0,liu2==0,liu5==0},{\[Lambda]0,\[Lambda]1,\[Lambda]2}];
l0b=\[Lambda]0 /.lsg[[1]];
l1b=\[Lambda]1 /.lsg[[1]];
l2b=\[Lambda]2 /.lsg[[1]];
l1b=\[Lambda]1/.lsg[[1]];
\[Lambda]0=l0b
\[Lambda]1=l1b
\[Lambda]2=l2b
\end{verbatim} \textBlack
Now one can insert the Lagrange parameters into the Liu equations and one gets the restrictions to the constitutive equations. The residual inequality determines the entropy production density.

\newpage
\bibliography{rel-therm}
\end{document}